\documentclass{caosp306}

\usepackage{graphicx}

\usepackage{natbib}
\articleNo{B25}
\pubyear{2019}
\volume{49}
\volnumber{2}
\firstpage{287}
\received{October 31, 2018}
\accepted{January 21, 2019}

\def\BibTeX{{\rm B\kern-.05em{\sc i\kern-.025em b}\kern-.08em
             T\kern-.1667em\lower.7ex\hbox{E}\kern-.125emX}}

\begin{document}

%
\hauthor{B.\,Seli, L.\,Kriskovics and K.\,Vida}

\title{Deriving photospheric parameters and elemental abundances for a sample of stars showing the FIP effect}

\author{
	B.\,Seli \inst{1,2}
      \and
    L.\,Kriskovics \inst{1}
      \and
    K.\,Vida \inst{1}
       }

\institute{
			Konkoly Observatory, Research Centre for Astronomy and Earth Sciences, Hungarian Academy of Sciences,
			1121 Budapest, Konkoly Thege Mikl\'os \'ut 15-17, Hungary \\
		  \and
			E\"otv\"os University, Department of Astronomy,
			1518 Budapest, Pf. 32, Hungary \\
           }

\date{November 1, 2018}

\maketitle

\sloppy
\begin{abstract}
One puzzling question in solar physics is the difference between elemental abundances in the photosphere and the corona. Elements with low first ionization potential (FIP) can be overabundant in the corona compared to the photosphere under certain circumstances. The same phenomenon has been observed on a handful of stars, while a few of them show the inverse effect. But not all the stars in the original sample had precise photospheric abundances derived from optical spectra, so for some the solar values were adopted. In this work we make homogeneous abundance measurements from optical spectroscopy.

We collected spectra of 16 stars showing the FIP effect with the 1-m RCC telescope of Konkoly Observatory, with resolution of $\lambda / \Delta \lambda \sim 21\,000$. We determine the fundamental astrophysical parameters ($T_\mathrm{eff}$, $\log g$, $[M/H]$, $\xi_\mathrm{mic}$, $v \sin i$) and individual elemental abundances with the SME spectral synthesis code using MARCS2012 model atmosphere and spectral line parameters from the Vienna Atomic Line Database (VALD).

\keywords{Stars: abundances -- Stars: atmospheres -- Stars: fundamental parameters -- Techniques: spectroscopic}
\end{abstract}

\section{Introduction}
\label{intr}
When working on X-ray spectra, solar physicists found a discrepancy between
the abundances of several elements compared to the known photospheric values.
In the solar corona, elements with low first ionization potential (FIP) are
enhanced by approximately a factor of 4 \citep[and references therein]{2015LRSP...12....2L}.
This phenomenon -- the FIP effect -- was later observed on a handful of stars.
The magnitude of the effect shows a spectral type dependence, for stars cooler than K5
the inverse FIP effect was observed, where the low FIP elements are depleted in the
corona.

However, for some of these stars the photospheric composition is unknown or the
available data were collected from several different sources. Substituting the
stellar elemental composition with the solar abundance pattern makes it impossible
to determine whether the coronal abundances are caused by the FIP effect or if those
elements are just over/underabundant in that particular star.
In this work we present new fundamental parameters and elemental abundances
derived from new homogeneous optical measurements for stars that are known
to show the FIP effect.

\section{Data}
We selected our target stars showing the FIP effect from Table 2 in \cite{2015LRSP...12....2L}. We excluded objects that are
not visible from Hungary, as well as faint targets (fainter than $V \sim 8^m$) to
ensure sufficient S/N. Our final observed sample consists of 16 main-sequence stars with
spectral types ranging from F6 to M3. The full list can be seen in Table \ref{sample}.

Observations were made with the 1-m RCC telescope of Konkoly Observatory,
equipped with an echelle spectrograph with $\lambda / \Delta \lambda \sim 21\,000$ mean resolution.
The observations were carried out in March, June, August and November 2017. The quality  of most
spectra is sufficient for spectral synthesis, while for the fainter stars we
have to combine spectra collected on the same night to ensure a high enough S/N.

\begin{table}[t]
\small
\begin{center}
\caption{The observed sample. Mean S/N was calculated with the DER\_SNR algorithm \citep{2008ASPC..394..505S}.}
\label{sample}
\begin{tabular}{lcc|lcc}
\hline\hline
star  & number of & S/N & star  & number of & S/N\\
  & spectra & & & spectra &\\
\hline
EK Dra	& 17  & 77 &    $\beta$ Com &	5 & 140 \\
EQ Peg A	& 24 & 39 & $\epsilon$ Eri &	42 & 100 \\
EV Lac &	23 & 23 &   $\kappa$ Cet &	20 & 113 \\
GJ 338 A &	31 & 69 &   $\xi$ Boo A &	5 & 124 \\
GJ 338 B &	27 & 66 &   $\xi$ Boo B &	5 & 58 \\
Sun &	3 & 99 &        $\pi^1$ UMa &	6 & 143 \\
70 Oph A &	5  & 122 &   $\pi^3$ Ori &	40 & 151 \\
70 Oph B &	15  & 80 &  $\chi^1$ Ori &	35 & 105\\
\hline\hline
\end{tabular}
\end{center}
\end{table}

Data reduction was carried out with the standard IRAF tasks. A ThAr spectral
lamp was used for wavelength calibration.
28 echelle orders were extracted from each image uniformly, but we restrict our
analysis to the 5000--7000\,\angstrom \,wavelength range, because below 5000\,\angstrom \,
the S/N gradually decreases while the region after 7000\,\angstrom \,is dominated
by telluric absorption lines.

\section{Spectral synthesis}
We used the Spectroscopy Made Easy (SME) code \citep{1996A&AS..118..595V} with MARCS2012 model atmosphere to calculate
the necessary parameters from the continuum normalized spectra. We downloaded spectral
line data from the Vienna Atomic Line Database \citep[VALD;][]{1995A&AS..112..525P} using the ``extract stellar''
option.

Before fitting the individual elemental abundances, the fundamental parameters of the
stars are needed. These are effective temperature ($T_\mathrm{eff}$), surface gravity ($\log g$), metallicity ([M/H]),
microturbulence ($\xi_\mathrm{mic}$) and projected rotational velocity ($v \sin i$).
Another necessary parameter is macroturbulent velocity ($\xi_\mathrm{mac}$),
but since it is hard to disentangle the contribution of $\xi_\mathrm{mac}$ from the other
line broadening effects, we chose to apply the following empirical relation from \cite{2005ApJS..159..141V}
rather than fitting $\xi_\mathrm{mac}$:
\begin{equation}
\xi_\mathrm{mac} = 3.5 + (T_\mathrm{eff} - 5777) / 650
\end{equation}
The following initial parameters were used: $\log g = 4.5$\,dex (since all stars in the sample are dwarfs),
$[M/H]=0$\,dex, $\xi_\mathrm{mic}=1\,{\rm km\,s}^{-1}$, $v \sin i = 20\,{\rm km\,s}^{-1}$
and $T_\mathrm{eff}$ inferred from spectral type. In general, no reasonable fit can be achieved if we iterate
all parameters at once, so we fit them in the following order:
first $\xi_\mathrm{mic}$ and $v \sin i$ simultaneously, then $T_\mathrm{eff}$, then $[M/H]$ and $\xi_\mathrm{mic}$.
After this step we fit $\log g$ with a line list containing only the Na D and Mg b lines, since their
strong line wings are more sensitive to gravity. Then we proceed by downloading a new line list from VALD
with the parameters derived so far. With this we fit $T_\mathrm{eff}$ once again, then $[M/H]$.
The results can be seen in Table \ref{results}, and Figure \ref{pi3ori} shows an example of the observed and synthetic spectrum of $\pi^3$\,Ori.

\begin{figure}[h]
\label{pi3ori}
\centerline{\includegraphics[width=0.75\textwidth,clip=]{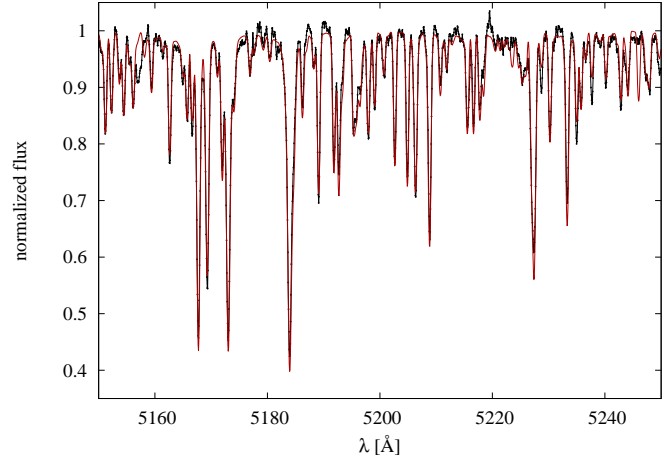}}
\caption{Observed (black) and synthetic (red) spectrum of $\pi^3$\,Ori near the Mg triplet.}
\label{pi3ori}
\end{figure}

After these steps we fit the individual elemental abundances (along with $[M/H]$), namely C, Na, Mg, Al, Si, S, Ca, Ti, Mn, Fe, Ni and Ba. Some interesting elements whose coronal abundance can be derived from X-ray spectra have no transitions in the observed wavelength range, so their abundances cannot be determined. The result can be seen in Figure \ref{hist_abund} for 8 stars from the sample. While most elements show no
deviation from the solar scale, there is a clear Ba enhancement for all these stars. According to \cite{2009ApJ...693L..31D} young stars tend to have higher Ba abundance, based on empirical results. All stars appearing in Figure \ref{hist_abund} are younger than the Sun with the oldest one being $\beta$ Com with age of $\sim$2.5\,Gyr, which explains the observed overabundance.
To check if any of these 8 stars could be Ba dwarfs we fit other s-process elements, although with larger uncertainties. It turns out that the abundances of La and Ce are high, while the Y and Zr content is approximately solar, which disproves the idea.
Al abundances also seem to be higher than solar, but since there are only a few and relatively weak Al lines to fit, it is likely not a real physical effect.

\subsection{Error estimation}
SME is robust enough to give almost identical results for all spectra collected from the same star on the same night.
This means that the standard deviations calculated from multiple observations are small (e.g. 5\,K in $T_\mathrm{eff}$)
and can only be used for consistency check. So the uncertainties of the derived parameters have to be obtained by different means, for example by seeing how much each parameter can be altered before it affects the determination of the other parameters during the fit.
The expected average uncertainties are 50\,K in $T_\mathrm{eff}$, 0.1\,dex in $\log g$, 0.1\,dex in $[M/H]$, 0.3\,km\,s$^{-1}$ in
$\xi_{\mathrm{mic}}$ and 3\,km\,s$^{-1}$ in $v \sin i$.

After comparing our results with available literature data, it seems that our $\log g$ values are usually lower
by $\sim$0.1\,dex. However that should have little effect on the derived abundances. Running the abundance fitting procedure
with 0.2\,dex difference in $\log g$ modifies the final abundances by $\sim$0.08\,dex. 200\,K change in $T_\mathrm{eff}$
results in $\sim$0.07\,dex difference, while 0.6\,${\rm km\,s}^{-1}$ change in $\xi_{\mathrm{mic}}$ gives $\sim$0.05\,dex.

\section{Conclusion}
It appears that a metre-class telescope equipped with a mid-high resolution spectrograph is enough to determine
elemental abundances for bright enough stars with the spectral synthesis method. We have collected spectra of 16 stars that show the FIP effect, and carried out the abundance analysis for 8 of them. The remaining stars are the fainter ones with noisier spectra, so for them the spectral synthesis will be more challenging. In the future we also plan to gather the available X-ray abundances to recalculate the FIP bias for these stars.

\begin{table}[t]
\small
\begin{center}
\caption{Fundamental parameters derived from the spectra.}
\label{results}
\begin{tabular}{cccccc}
\hline\hline
star  & $T_\mathrm{eff}$ & $\log g$ & $[M/H]$ & $\xi_\mathrm{mic}$ & $v \sin i$\\
 & [K] & [dex] & [dex] & [km\,s$^{-1}$] & [km\,s$^{-1}$]\\
\hline
EK Dra & 5780  & 4.46 & $-0.06$  & 1.32 & 21.2\\
$\beta$ Com & 5980  & 4.37 & $-0.09$  & 0.92 & 13.3\\
$\epsilon$ Eri & 5150  & 4.32 & $-0.07$  & 0.91 & 12.8\\
$\kappa$ Cet & 5780  & 4.39 & $-0.01$  & 0.83 & 12.5\\
$\xi$ Boo A & 5670  & 4.56 & $-0.16$  & 1.32 & 14.1\\
$\pi^1$ UMa & 5880  & 4.39 & $-0.18$  & 0.98 & 16.9\\
$\pi^3$ Ori & 6320  & 4.37 & $-0.12$  & 1.06 & 20.0\\
$\chi^1$ Ori & 5940  & 4.44 & $-0.10$  & 0.54 & 17.2\\
\hline\hline
\end{tabular}
\end{center}
\end{table}

\begin{figure}
\label{hist_abund}
\centerline{\includegraphics[width=1.0\textwidth,clip=]{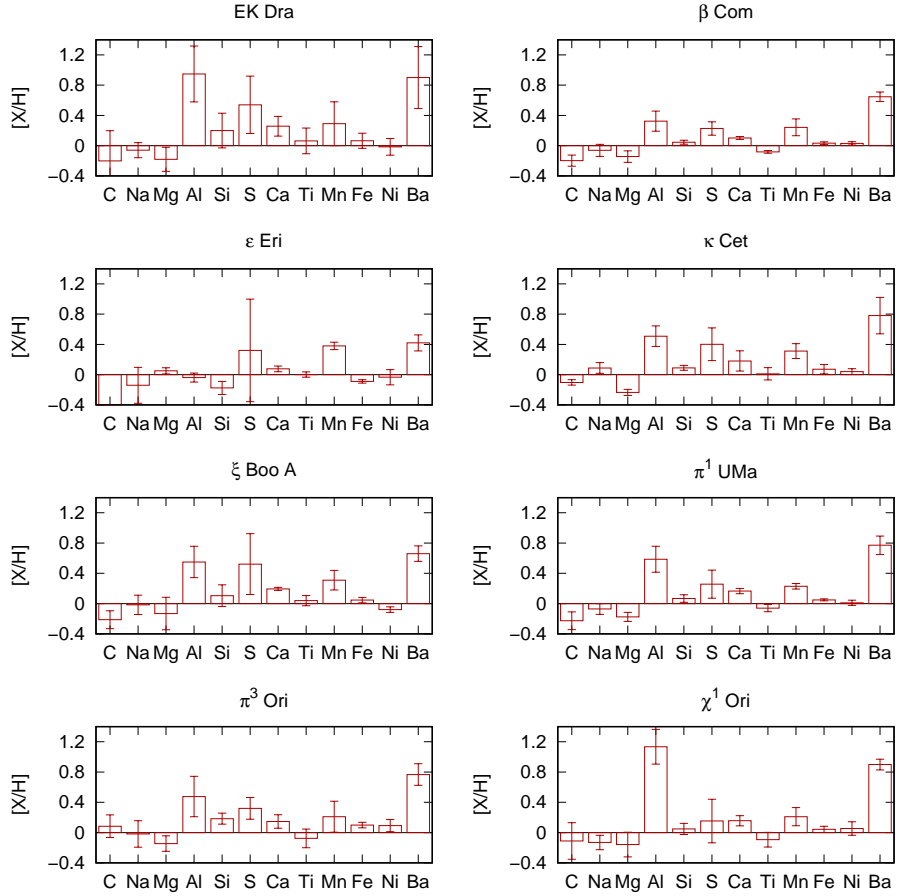}}
\caption{Elemental abundances relative to the solar values (derived in this work). Error bars shown are multiplied by 10 for illustration purposes.}
\label{hist_abund}
\end{figure}

\acknowledgements
Authors are grateful to Konkoly Observatory, Hungary, for hosting two workshops on Elemental Composition in Solar and Stellar Atmospheres (IFIPWS-1, 13--15 Feb, 2017 and IFIPWS-2, 27 Feb--1 Mar, 2018) and acknowledge the financial support from the Hungarian Academy of Sciences under grant NKSZ 2018\_2.
The authors acknowledge the Hungarian National Research, Development and Innovation Office
grant OTKA K-113117 and the Lend\"ulet grant LP2012-31 of the Hungarian Academy of Sciences.
KV is supported by the Bolyai J\'anos Research Scholarship of the Hungarian Academy of Sciences.
This work has made use of the VALD database, operated at Uppsala University, the Institute of Astronomy RAS in Moscow, and the University of Vienna.
Authors are grateful to Borb\'ala Cseh for her helpful suggestions related to Ba stars.

\bibliography{demo_caosp306}

\end{document}